# SE Research is a Complex Ecosystem

Isolated Fixes Keep Failing—and Systems Thinking Shows Why

Mary Shaw
Software & Societal Systems Dept
Carnegie Mellon University
Pittsburgh PA USA
mary.shaw@cs.cmu.edu

Mary Lou Maher
School of Computer Science
University of Syndney
Sydney NSW Australia
Marylou.maher@sydney-edu.au

Keith Webster
University Libraries
Carnegie Mellon University
Pittsburgh PA USA
kwebster@andrew.cmu.edu

## ABSTRACT

The software engineering research community is productive, yet it faces a constellation of challenges: swamped review processes, metric-driven incentives, distorted publication practices, and increasing pressures from AI, scale, and outright scams. These issues are often treated in isolation, yet they arise from deep structural dynamics within the research ecosystem itself and distract us from the larger role of research in society. Meaningful progress requires a holistic system-level view. We sketch such a framework drawing on ideas from complex systems, ecosystems, and theory of change. Reframing SE's challenges through this lens reveals non-linear feedback loops that sustain current dysfunctions, and it helps to identify leverage points for reform. These are less a matter of isolated fixes and more a matter of exploring coordinated sets of fixes that operate across the SE ecosystem

## CCS CONCEPTS

• Software and its engineering

## 1. State of SE research

The software engineering (SE) research community is fundamentally healthy and is producing a steady stream of important results. Nevertheless, like other disciplines, it is facing a growing set of well-documented challenges in the development, maintenance, and curation of the scholarly record. The peer review process is overloaded, and publication volumes are escalating faster than quality control mechanisms can scale. Conferences and journals struggle to balance rigor with timely publication, and new approaches are harder to publish than incremental results. Career progression often targets bibliometric indicators rather than substantive intellectual contribution. Institutional priorities are often driven by outside forces to prioritize those bibliometrics. These forces threaten trust in the scholarly record itself. Early-career researchers often pursue approaches that leave little time for reflection, synthesis, or risk-taking. These problems are widely recognized and frequently discussed, yet proposed solutions have had limited success.

Crucially, these problems do not arise independently, nor can they be solved in isolation. Efforts to optimize one part of the system, such as increasing selectivity or expanding review capacity, often trigger compensating behaviors elsewhere, such as resubmission cascades and reviewer fatigue.

The challenges facing the SE research community are intertwined rather than isolated. The community functions as an ecosystem of components with varied responsibilities (see section 3) whose collective behavior is driven by their interdependencies. Understanding why interventions fail, and how meaningful improvement might be achieved, requires treating SE research as a complex system with rich interdependencies. We adopt this perspective to examine how core research functions are currently discharged and where opportunities for change may lie.

## 2. Core responsibilities of a research community

Research is embedded in society. Since 1945, research has been fueled by a social contract under which industrial innovation is powered by a flow of basic research and talent that is supported by government funding [2]. This social contract entails a core set of responsibilities that the research community must fulfill. To understand the SE research community as a complex ecosystem we must understand the breadth and depth of these responsibilities, summarized as:

1. *Knowledge Creation:* A research community must generate new knowledge through sustained inquiry, experimentation, and synthesis that advances rigorous understanding rather than merely increasing output or pursuing novelty.
2. *Validation:* The community must assess the quality, rigor, and reliability of knowledge claims through credible, scalable mechanisms that distinguish sound work from flawed or fraudulent results.

3. *Dissemination:* The community must communicate research outputs effectively and equitably, ensuring that knowledge reaches appropriate audiences without distortion by prestige or metrics.
4. *Preservation & Stewardship:* The community must preserve and steward the scholarly record, including data, methods, and context, so that knowledge remains accessible, interpretable, and reusable over time.
5. *Training & Inclusion:* The community must maintain a rich pipeline of researchers by instilling knowledge, values, skills, and norms while reducing structural barriers to participation.
6. *Recognition & Reward:* The community must allocate credit, status, and resources in ways that motivate and reinforce rigor, collaboration, and long-term contribution rather than short-term visibility or volume.
7. *Societal Integration:* The community must connect research to societal needs and impacts, maintain public trust, and ensure that knowledge contributes to broader human and institutional outcomes.

The ability of the research community to fulfill these requirements, and hence the current challenges facing the community, are emergent behaviors influenced by internal and external factors. Internal factors are those that the community can control, for example the effect of bibliometric-driven career advancement, which leads to reviewing overload. External factors are those that drive change from outside the community, for example the availability of Gen AI to generate and review papers. To understand and mitigate the current challenges, we must identify the components in the ecosystems and the ways their interests and actions interact with respect to these responsibilities.

## 3. SE research as a complex ecosystem

The core responsibilities identified in Section 2 are carried out by a variety of components–institutions and individuals. To understand the emergent behavior of the interactions among these components, we view them as a complex ecosystem.

A *complex system* is one composed of many interacting parts where the relationships between those parts give rise to collective behaviors that are difficult to predict from the properties of the individual components alone. Complex systems have been used to model natural and engineered systems as well as human interaction, with a common set of principles [6]. For present purposes, we take the components of the SE research system to be individual researchers, research institutions, funding agencies, conferences, publishers, archival literature, gray literature (blogs, preprints, social media, white papers, etc), other organizations (professional societies, ranking organizations, governments), and the companies and consumers that apply the research in practice. Characterizing the SE research community as a complex system acknowledges the non-linearity of the system: small changes can lead to large effects through feedback loops in which outputs route back as inputs (for example, research findings influence funding priorities). Emergent higher level behaviors arise from lower level interactions (eg reviewing overload is a result of expectations for reward and evaluation as well as increasing numbers of early career actors).

An *ecosystem* is defined through the roles, interdependence, and environment of its components. Ecosystem models, originally developed to model the natural environment [4], are also used to model communities, for example the STEM ecosystem models [12]. Acting as an ecosystem, the components of the SE research community are affected by various agents (biotic components) and their environment (abiotic components). Recent significant environment changes include the rise of Generative AI, the rise and fall of funding for basic research, and the economic incentives for malicious actors such predatory journals. The ecosystem model highlights interdependence, niches and diversity, flows, and the critical effect of environment or context:

- *Interdependence:* All components—researchers, educators, funders, publishers, and the public—rely on each other for the system's health and productivity.
- *Niches and Diversity:* Different components have distinct roles (e.g., basic vs. applied research, teaching vs. R&D); the diversity of these roles is critical for resilience and progress.
- *Flows (Resources):* Essential resources (like funding, knowledge, talent, and data) move through the system (analogous to energy and nutrient flow).
- *Environment/Context:* The SE research community is embedded in a broader socio-political and economic landscape (the "physical environment") that shapes its survival and evolution.

A *theory of change* can guide a plan for addressing the challenges facing the SE research community, while acknowledging the complex and emergent behaviors typical in a complex ecosystem. Developing a robust theory of change can be challenging, but it is an essential precursor for increasing the potential for effecting positive changes in a complex ecosystem [9]. One approach in a theory of change is to identify interventions and intended outcomes, and overlay those on the components of the systems model to enable an analysis of the effect and impact of the interventions on the responsibilities and values of the research community. For example, an intervention that changes the evaluation of early career researchers to focus on quality rather than quantity of publications could have the intended outcome of reducing the number of papers each early career researcher submits to high-profile conferences. But that alone is insufficient if other forces such as funding, ranking, and hiring, persist in emphasizing quantity. Analyzing the effects involves identifying which components are affected and how the rules of interaction shift behaviors.

Each of the components in the ecosystem contributes to multiple core responsibilities, either acting deliberately to fulfill the responsibility or providing suitable resources. A change in one component may have effects across multiple components and their collective ability to fulfill some responsibilities, leading to complex interdependencies. Further, the system's behavior is driven not by money alone, but by competing value flows such as reputation, attention, time, and institutional status, which interact in powerful feedback loops. Table 1 maps the components of the ecosystem (the rows) to the core responsibilities of a research community (the columns), showing a number of the ways the components act or provide resources to fulfill the responsibilities.

Table 1 highlights three features of the SE research ecosystem that are easy to miss. First, no core responsibility is owned by a single component; all depend on coordination across multiple components of the system. Second, several responsibilities are heavily concentrated in conferences and publishers, creating bottlenecks when volume increases. Third, preservation, synthesis, and societal integration are comparatively weakly anchored, often assumed rather than explicitly supported. These patterns help explain why local interventions frequently displace problems rather than resolving them. Changing incentives for one component affects multiple responsibilities and actors.

|  | Knowledge Creation | Validation | Dissemination | Preservation & Stewardship | Training & Inclusion | Recognition & Reward | Societal Integration |
|---|---|---|---|---|---|---|---|
| *Researchers* | **do the work** | bear the responsibility | bear the responsibility |  | **teach & mentor** |  | **mentor** |
| *Research institutions* | host facilities |  |  | host repositories |  | **hire & promote** | mentor |
| *Funding agencies* | pay for research |  |  |  | Require training | fund | expect impact |
| *Conferences* | offer interaction venue | **gatekeep via review** | **publish results** | add to archive |  | feed bibliometrics; give awards | **facilitate collaboration** |
| *Publishers* | host literature | **gatekeep via review** | **publish results** | add to archive |  | feed bibliometrics |  |
| *Archival literature* | host reference materials |  | index & give access | **preserve permanently** |  |  |  |
| *Gray literature* | host other works | provide informal evidence | give diffuse access |  |  | give viral recognition | build community |
| *Other orgs* |  | set standards |  |  | **define curricula** | **Give awards; set expectations** |  |
| *Busines, consumers* | **motivate research** | adopt innovations | incorporate in products |  | hire graduates | funding | **market application** |

**Table 1. The ways that SE research ecosystem components (rows) contribute to core responsibilities (columns). Bold entries indicate principal roles.**

## 4. Complex dependencies among publication forces

With this context as a lens, we can examine some of the challenges facing the SE research community. We specifically consider the part of the ecosystem that supports the dissemination responsibility: one of the most severe current pain points is the flood of reviewing tasks; further, paper mills and generative AI threaten to make this even worse. We first consider the current tsunami of papers and the ways it threatens the social contract underlying research. We then turn to other possible models for publication.

To scale the problem, the community survey for this track [11] reports that over 18% of respondents were (co-)author on over 20 SE conference/journal papers in the past three years, of whom almost 7% (co-)authored over 40. Over half of the 280 respondents report that some aspect of publication is not working well, especially reviewing overload, emphasis on quantity over quality, and relevance to industry. Over 40% named some aspect of publication as the one thing they would change, and over 40% named reviewing or publishing as their greatest source of stress.

### 4.1 Understanding the dynamics of the paper tsunami
We use the concepts of a theory of change and complex ecosystem lens to examine a set of related challenges that arise from the current fixation on rapid publication of many papers. These are subject to an external constraint, the social contract that provides research support and expects this to feed innovation in practice. We note some of the ways the paper tsunami interferes with fulfilling our obligations. Our examples draw heavily on the pre-survey for this track [11].

We can identify many forces on paper submission of the form <force on component1> drives <component2> to <action affecting component3>. These can be chained to see cascading effects, including cycles that become positive feedback loops[1]. To see how these challenges interact to power the tsunami, consider just a small sample chain of forces and the actions that respond to them:

university ranking schemes drive university promotion policies to count selective publication
which drives faculty to publish more in selective conferences and
the role of selectivity drives the conferences to be even more selective
which drives up the number of papers submitted and resubmitted
which drives up the reviewing load nonlinearly
which means faculty have less time to evaluate research quality instead of just counting papers (and grants) and
(may) influence the ranking organizations to raise their standards, and
the publication flood means new PhDs need double-digit papers in good places to get interviews
which enables employers to use publication counts to select applicants
which drives up the number of paper submitted and resubmitted
which means researchers are demanding more attention from other researchers (and page charges add up)
which means researchers are too busy writing papers to synthesize their results into larger theories
which means new PhDs are trained to believe that this is good science

---
[1] Positive feedback amplifies effects, often exponentially; here it's undesirable.

CRA attempted to break this cycle by calling for evaluation based on 3-5 papers [3]. Simply addressing evaluation of researchers did not address the values held by other components of the ecosystem that feed the pursuit of paper counts and has not, in isolation, been successful.

The values, decisions, and actions of many independent components interact to reinforce the cascade. We see in particular a pernicious codependence among rating agencies seeking "objective" measures, conferences striving for "highly selective" status via artificially high rejection rates, universities seeking stature by using bibliometrics as a proxy for quality, and researchers pursuing positions and reputation in the face of these pressures.

We can identify a number of specific challenges in this behavior, along with some candidate interventions. However, none of the individual interventions can work in isolation – the forces that create publication pressure must be relieved for many components, not just one.

- *Reviewing challenge:* This is the loop everyone seems to see first: pressure for publication counts creates a reviewing crisis. *Candidate intervention:* Restore the social norm that you owe as much reviewing as you get, plus more (not everyone getting reviews should be doing reviews). *Likely failed intervention:* IJCAI plan to charge $100 for all but first submission [5].
- *Conference challenge*: Conferences provide opportunities for interaction and socialization, but it has become common that support to attend a conference depends on presenting a paper there. Also, they're too expensive, which undercuts equity of access – a principal (social) function of conferences – and it amplifies the paper-production crisis. *Candidate interventions*: Reset norms to encourage attendance without papers at a reasonable number of conferences; change conference formats; make more regional conferences (e.g., within driving distance); make hybrid conferences actually work
- *Job-hunting challenge*: Not only does it "take 10 good papers" for a new PhD to get a good academic interview, but for industrial positions as well. This looks like a classic snowballing response to a positive-feedback loop—if applicants think they are ranked by number of papers, then the number of papers escalates. Promotions may be similarly affected. *Candidate interventions*: CRA tried to shift from quantity to quality [3], but this intervention calls for a shift in values across multiple components. Hiring organizations could publicize—and observe—more appropriate criteria. John Hopcroft suggested that evaluations should be based on three papers selected at random from your CV. Brian Randell suggests that each researcher be given a lifetime "Academic Ration" that allows them to submit for publication a very restricted number of "research articles" per decade, or an unlimited number of rigorously reviewed "academic survey papers".
- *Broken-science challenge:* Good science requires more than a flood of individual discrete results. The incentive to publish lots of papers diverts effort from synthesizing those results into more comprehensive theories and integrating them to be useful in application. In other words, the rest of the research enterprise is neglected. Even worse, we are training new scientists to believe this is normal; they may not even realize how narrow a slice of good science this is. *Candidate interventions*: Whatever will break the "more papers" fixation; increasing article processing charges could be a deterrent, but its effect would be inequitable.
- *Paper mill and AI challenge*: They pollute the research stream so you can't find real results, creating further overload. *Candidate intervention:* For paper mills, if single incremental papers were not the primary currency of recognition, would they lose their market?

Each of these challenges interferes in one way or another with fulfilling our core responsibilities as a research community. For example, the paper tsunami overwhelms our capacity to do good reviews, which causes us to fall short of fulfilling our responsibility of validation. Table 2 shows which of the core responsibilities are short-changed in each of these challenges.

| | Knowledge Creation | Validation | Dissemination | Preservation | Training, Inclusion | Recognition | Societal Integration |
|---|---|---|---|---|---|---|---|
| Reviewing | | X | | | X | | |
| Conference | | | X | | X | | |
| Job hunting | | X | | | | X | |
| Broken science | X | X | | | X | X | X |
| Paper mills/AI | X | X | X | X | | X | X |

**Table 2. How current challenges (rows) make the CS research community fall short of satisfying its responsibilities (columns)**

## 4.2 Reassessing the publication landscape

The forces that drive the paper tsunami have made the individual conference paper the dominant unit of scholarly output. This has crowded out other forms of research, and it risks training new researchers to equate paper output with scientific contribution and hence to prioritize small results over cumulative science. This works to the detriment of our responsibility for societal integration and leads to a literature in which all the bits of progress are available, but they lack a strong synthesis. It is as if you asked GitHub for the current version of an app and it gave you a set of pull requests to integrate for yourself.

In other disciplines, systematic reviews or survey papers have distinguished status. SE has ACM *Computing Surveys* as an archival venue for these – but many of our colleagues treat these as secondary for reputation. An indication that we don't take them seriously is that the ACM and IEEE SE journals published over 600 papers in 2025 (conferences published hundreds more) but *Computing Surveys* published only 5 SE papers and 7 others arguably related. At the other end of the spectrum of rigor, we deposit unrefereed papers on preprint servers, and these wind up in reference lists without indications of whether they have passed rigorous review (as do blog posts and random web sites).

There have been some countervailing arguments: Brooks distinguished three classes of results: findings, observations, and rules-of-thumb, each appropriately evaluated. [1]. Evidence-based disciplines such as medicine recognize several levels of evidence and aggregate them to reach strength of recommendation conclusions. SE could adopt similar approaches. [7][8].

The effect of the paper tsunami is that the SE publication landscape is lacking at both the higher level that synthesizes individual results and at the more informal level of the gray literature, which lacks conventions for recognizing the strength of results.

Rather than identify individual interventions for specific components of the system or for each challenge, a theory of change is enacted across the components of the systems and calls for collective interventions. A change in the values associated with SE publications should be applied as a series of interventions that cut across the components of the ecosystem. We propose to normalize the description and categorization of research outputs across all components of the system. The categories might be, for example:

- *Syntheses:* properly executed and reviewed meta-analyses, systematic literature reviews, rigorous surveys
- *Primary research:* validated archival papers in fully reviewed journals and main research tracks of certain conferences
- *Observational studies:* surveys and experience reports based on sound qualitative methods
- *Preliminary results:* reviewed papers in workshops, auxiliary conference tracks (experience reports, new ideas)
- *Artifacts:* publicly available data sets, software, benchmarks
- *Gray literature:* preprints, technical reports, essays
- *Social media:* blogs and other unvalidated opinions

This intervention must be coordinated with the way we reward researchers: the value associated with dissemination is not simply a function of the sheer number of publications but is also related to the distribution of publications across categories, especially higher categories.

How, then, would we actually transition to a state in which a wide range of result types were appropriately honored? First, we would need interventions that break the stranglehold of the current fixation on individual results as discussed in Section 4.1. This would require consensus on result types, their stature, and criteria for evaluating each. In particular, it could de-emphasize treating most conference papers as archival, thereby changing their status and their need for rigorous review. This would require refactoring the current review and archiving process to match the stature of the result type with rigor of review and the type of preservation to provide (we certainly want workshop papers to be available for a few years, but do they really need to be permanently archived?). We would need encouragement for doing more syntheses: perhaps NSF could incorporate this in expectations of proposals. This would also require rethinking standards for hiring and promotion.

## 5. Need for holistic view of SE research ecosystem

The SE research community is a complex ecosystem with tightly coupled interacting components. Changes in one part of the ecosystem usually propagate through the system, often with unexpected results. We have used a particular example, the publication tsunami, to illustrate this coupling, and we have identified a number of potential interventions. More significantly, we have argued that these interventions shouldn't be applied in isolation, but coordinated with a consensus on a change in how we categorize and value publications.

Our larger point is that the SE community should invest in a better understanding of our ecosystem and its complexity. We propose a holistic framing that integrates complex system and ecosystem views. We show how these lenses illuminate some of our problems and argue that this kind of analysis is crucial for progress, that we must make deliberate choices of complementary sets of interventions.

This is particularly timely, because external threats such as paper mills and AI slop are rapidly increasing [10]. We won't be able to fend them off with isolated interventions. Effective defenses against these and other malicious agents will require consensus on a model of the complex ecosystem of SE such as we propose here.

A systems view is essential to controlling our destiny in the face of rapidly-changing technologies.

## ACKNOWLEDGMENTS

Mary Shaw is supported by the A.J. Perlis Chair of Computer Science at Carnegie Mellon University. Mary Lou Maher is supported by the Sydney Horizons initiative at the University of Sydney. Keith Webster is supported by the Helen and Henry Posner Jr. Dean's Chair at Carnegie Mellon University.